\documentstyle{article}
 \newcommand{\beq}{\begin{equation}}
 \newcommand{\eeq}{\end{equation}}
 \newcommand{\beqa}{\begin{eqnarray}}
 \newcommand{\eeqa}{\end{eqnarray}}
 \newcommand{\bea}{\begin{eqnarray}}
 \newcommand{\eea}{\end{eqnarray}}

 \newcommand   {\Hmc}     {H_{\mbox{\tiny MC}}}

 \newcommand   {\bl}      {\sigma^{(s)}}
 \newcommand   {\fl}      {\psi^{(s)}}
 \newcommand   {\ff}      {\psi^{(4)}}

 \newcommand   {\ev}[1]   {\langle #1\rangle}

 \input{psfig}
 
\begin{document}
 \begin{titlepage}
 \begin{flushright}
 LU-TP 96-12\\
Revised version\\
\today
 \end{flushright}
 \vspace{0.8in}
 \LARGE
 \begin{center}
 {\bf Identification of Amino Acid Sequences with \\ 
Good Folding Properties in an Off-Lattice Model}\\

 \vspace{.3in}
 \large
 Anders Irb\"ack\footnote{irback@thep.lu.se}, 
 Carsten Peterson\footnote{carsten@thep.lu.se} and 
 Frank Potthast\footnote{frank@thep.lu.se}\\
 \vspace{0.10in}
Complex Systems Group,  Department of Theoretical Physics\\ 
University of Lund,  S\"{o}lvegatan 14A,  S-223 62 Lund, Sweden \\
 \vspace{0.3in}
 
{\it Physical Review} {\bf E 55}, 860 (1997) 
 
 \end{center}
 \vspace{0.3in}
 \normalsize

Abstract:

Folding properties of a two-dimensional toy protein model containing only 
two amino-acid types,  hydrophobic and hydrophilic respectively, are 
analyzed. An efficient Monte Carlo procedure is employed to ensure that 
the ground states  are found. The thermodynamic properties are found to be
strongly sequence dependent in contrast to the kinetic ones. Hence, criteria 
for good folders are defined entirely in terms of thermodynamic fluctuations.  With these criteria sequence patterns that fold well are isolated. For 300 chains 
with 20 randomly chosen binary residues approximately 10\% meet these 
criteria. Also, an analysis is performed by means of statistical and artificial 
neural network methods  from which it is concluded that the folding properties  can be predicted to a certain degree given the binary numbers characterizing 
the sequences.

 \end{titlepage}

 \newpage

 \section{Introduction}

 The protein folding problem is not merely an engineering task -- given sequences 
 of amino acid residues, compute its 3D structure by minimizing an appropriately 
 chosen energy function.  Since for such models the energy landscape is often rugged, 
 the resulting 3D configurations may be hard to reach and may furthermore not be 
 thermodynamically stable.  It has therefore been argued that only those sequences 
 with ``nice'' energy landscapes have survived the evolution \cite{sali}. 

 A proper understanding of the thermodynamics and kinetics of protein folding 
 requires studies of simplified toy models where the conditions can be somewhat 
 controlled. For the choice of such models two major pathways exist. The
 currently most popular choice is lattice models with contact term   
 interactions, see e.g. Refs.~[1--4]. This approach has the advantage that 
 the ground states are known, but at 
 the same time it has the potential danger that the energy landscape contains artifacts 
 from the discrete description of space. Alternatively, one may use a continuum 
 model with simplified interactions, see e.g. Refs.~[5--7], 
 in which case substantial 
 simulations are needed to map out the ground states. On the other hand, in 
 this case properties of the energy landscape should be closer to those of the 
 real world. 

 The aim of this paper is twofold - to map out the folding properties of 
 the  two-dimensional
 continuum model of \cite{still1,irback}, hereafter denoted the $AB$ model, 
 and to analyze how the folding properties depend upon the sequences using 
 statistical and state-of-the-art regression methods.

 The folding properties of the $AB$ model are investigated 
 with respect to thermodynamics and kinetics given a set of thoroughly simulated 
 sequences.  In total 300 sequences of 20 hydrophobic 
 and hydrophilic residues ($+1$ and $-1$ respectively) are studied using an efficient
 dynamical-parameter algorithm (see Ref.~\cite{irback} and references therein). 
The thermodynamic properties are studied using the mean-square 
distance $\delta^2$ between 
different configurations. A low average value 
$\ev{\delta^2}$ signals that the chain exists in a state with well-defined structure.  
It turns out that $\ev{\delta^2}$ exhibits very strong sequence dependence in 
contrast to the kinetic properties.
Based on this we formulate criteria for good folders 
entirely based on the distribution of $\delta^2$. Using these criteria roughly 
10\% of the 300 generated and studied sequences survive as good folders. 

Next we pose the question of what characterizes the good folders in terms of 
sequence patterns. Rather than analyzing the ``bare'' binary sequences of 
hydrophobicity, we focus on effective variables like random walk representations, 
block fluctuations and the number of +1 embedded between two -1 . This  has the 
virtue that the analysis will capture long range correlations in addition to the 
local ones. We investigate how $\ev{\delta^2}$ depends upon these quantities. This is 
done using tools of varying sophistication -- covariance matrix and feedforward 
Artificial Neural Networks (ANN).  Using ANN we predict $\ev{\delta^2}$ given the 
sequence. With our limited data set the results look very promising. Indeed, the folding 
properties  strongly depend upon sequence patterns. These findings give further evidence 
of the non-randomness reported in Ref.~\cite{irback1}.  

 Hydrophobicity is widely believed to play a central role in the formation of 
 3D protein structures. In Ref.~\cite{irback1} the question of whether proteins 
 originate from random sequences of amino acids was addressed by means of a 
 statistical analysis in terms of blocked and random walk values formed by 
 binary hydrophobic assignments of the amino acids along the  protein chains. 
 The results, which were based upon  proteins in the SWISS-PROT data 
 base~\cite{swiss-prot},  convincingly  demonstrated that the amino acid sequences 
 in proteins differ from what is expected from random sequences in a statistical 
 significant way.  In Ref.~\cite{irback1} also preliminary results from the $AB$  
 model using the same data as in this work were subject to the same statistical 
 analysis. The interesting observation was made that the AB model sequences that fold 
well according to  low $\ev{\delta^2}$-value criteria exhibit similar deviations  from randomness as for  the functional proteins. The deviations from
 randomness can be interpreted as  originating from anticorrelations in terms of 
an Ising spin model for the  hydrophobicities. 

Our studies of $AB$ model are limited to two dimensions in order to be 
able to analyze many sequences within limited CPU resources. How realistic 
this approximation is, can of course be questioned. The system may be "stiffer" 
than a three dimensional one when it comes to swapping monomer positions.

 This paper is organized as follows. In Sec.~2 we briefly describe the 
 model and generation of sequences. The Monte Carlo method and what is 
 being measured can be found in Sec.~3.  The thermodynamics and kinetics of the 
 system are described in Sec.~4, whereas Sec.~5 contains our statistical and 
 ANN analysis. 
In Sec.~6 we briefly review the results from Ref.~\cite{irback1} comparing deviations 
from non-randomness in the two-dimensional $AB$  model with those of functional proteins. 
A brief summary can be found in Sec.~7.

 \section{The Model}

 \subsection{General Formulation}

 The $AB$ model consists of two kinds of monomers,  $A$ and $B$ respectively. 
 These are linked by rigid bonds of unit length to form linear chains living in 
 two dimensions. For an $N$-mer  the sequence of monomers is described 
 by the binary variables $\sigma_1,\ldots,\sigma_N$ and the configuration by the 
 angles $\theta_2,\ldots,\theta_{N-1}$, where $\theta_i$ denotes the bend 
 angle at site $i$ and is taken to satisfy $|\theta_i|\le\pi$. The energy function 
 is  given by 
 \beq
 E(\theta,\sigma) =  \sum_{i=2}^{N-1} E_1(\theta_i) + 
 \sum_{i=1}^{N-2}\sum_{j=i+2}^N E_2(r_{ij},\sigma_i,\sigma_j)
 \label{e}
 \eeq   
 where 
 \bea
 E_1(\theta_i)&=&{1\over 4}\,(1-\cos\theta_i) \nonumber \\
 E_2(r_{ij},\sigma_i,\sigma_j)&=&
 4(r_{ij}^{-12}-C(\sigma_i,\sigma_j)r_{ij}^{-6})
 \label{e12}
 \eea
 and $r_{ij}=r_{ij}(\theta_{i+1},\ldots,\theta_{j-1})$ denote the distances 
 between sites $i$ and $j$. The term $E_1(\theta_i)$ favors alignment of three 
successive sites; $i-1$, $i$ and $i+1$. The nonbonded interactions $E_2$ 
 are Lennard-Jones potentials with a species-dependent coefficient $C(\sigma_i,\sigma_j)$, 
 which is taken to be 1 for an $AA$ pair (strong attraction),  1/2 for a $BB$ pair 
 (weak attraction) and -1/2 for an $AB$ pair (repulsion). Consequently,  there is 
 an energetic preference for separation between the two kinds of monomers. In fact, 
 it was demonstrated in \cite{still1} that ground-state configurations tend to have 
 a core consisting mainly of $A$ monomers, which shows that $A$ and $B$ 
 monomers behave as hydrophobic and polar residues respectively. The behavior 
 of the model at finite temperature $T$ is defined by the partition function 
 \beq
 Z(T,\sigma)=\int 
 \biggl[\prod_{i=2}^{N-1}d\theta_i\biggr] 
 \exp(-E(\theta,\sigma)/T)\ .
 \label{z}
 \eeq

 \subsection{The Sequences}

In total 300 sequences were drawn randomly from the set of all distinguishable 
chains with 14 A and 6 B monomers. 
Our motivation for this somewhat arbitrary choice of A/B ratio is that  
there are thermodynamically stable structures at relatively high 
temperatures for this ratio~\cite{irback}. 
%This choice of A/B ratio is motivated by allowing for cores of A's surrounding by B's.  
This set contains 19980 sequences, 
whereas  the total number of sequences with the same composition is 38760.  
Among the  300 sequences 4 are symmetric.  The 300 distinguishable sequences can 
be taken as 300 independent sequences drawn from the distribution of all sequences with double weight for every asymmetric sequence.

 \section{Simulations}
 \subsection{Methods}
We have performed numerical simulations of both the thermodynamic and kinetic
behavior of the 300 randomly selected sequences. At low temperature 
the system is in a folded phase with high free-energy barriers, which 
makes conventional simulation methods very time-consuming. As in 
Ref.~\cite{irback}, we therefore employ the dynamical-parameter method
for the thermodynamic simulations. In this approach one tries to accelerate 
the simulation by letting some parameter of the model become a 
dynamical variable, which takes values ranging over a definite set.
In the present work we have taken the temperature as a dynamical 
parameter (``simulated tempering'' \cite{marinari}) which means that we 
simulate the joint probability distribution 
\beq
\label{Pdis}
P(\theta,k)\propto\exp(-g_k-E(\theta,\sigma)/T_k) \,
\eeq
where $T_k$, $k=1,\ldots,K$, are the allowed values of the temperature.  
The $g_k$'s are tunable parameters which must be chosen carefully
for each sequence. The determination of these parameters has been carried  
out by the same methods as in Ref.~\cite{irback}.    

The joint distribution $P(\theta,k)$ is simulated by using an ordinary 
Metropolis step~\cite{metropolis} in $k$ and a hybrid Monte Carlo 
update~\cite{duane} of $\theta$. The hybrid Monte Carlo update is based on
the evolution arising from the fictitious Hamiltonian
\beq
\Hmc(\pi,\theta)={1\over 2}\sum_{i=2}^{N-1}\pi_i^2 + E(\theta,\sigma)/T 
\label{ham}\eeq
where $\pi_i$ is an auxiliary momentum variable conjugate to $\theta_i$.
The first step in the update is to generate a new set of 
momenta $\pi_i$ from the equilibrium distribution
$P(\pi_i)\propto \exp(-\pi_i^2/2)$. Starting from these 
momenta and the old configuration, the system is evolved through 
a finite-step approximation of the equations of motion. 
The configuration generated in such a trajectory is finally 
subject to an accept-or-reject question, which removes errors due to the 
discretization of the equations of motion.
The hybrid Monte Carlo update has two tunable parameters, 
the step size $\epsilon$ and the number of steps  $n$ in each trajectory.

The dynamical-parameter method greatly improves the frequency of transitions 
between different free-energy valleys, as compared to plain hybrid Monte Carlo.
In Ref.~\cite{irback} a speed up factor of almost $10^3$ was observed
for system size $N=10$. For $N=20$ we expect the gain to be even larger.
  
In our simulations we used a set of $K=13$ allowed temperature values,
which were equidistant in $1/T$ and ranging from 0.15 to 0.60.
Each hybrid Monte Carlo trajectory was followed by one Metropolis
step in $k$. The step size parameter $\epsilon$ was taken to vary 
with $T$, from 0.0025 at $T=0.15$ to 0.005 at $T=0.6$, while $n=100$ was held 
fixed. For the typical sequence the average acceptance rate was 
around 95\% for the $\theta$ update and 65\% for the $k$ update. For each
sequence a total of 440000 update cycles were carried out, which requires 
around 4 CPU hours on a DEC Alpha 2000.

In order to study the kinetic behavior of the model, we have performed hybrid Monte 
Carlo simulations at different fixed values of $T$. Starting from random coils,
we study the rate of the subsequent relaxation process.
While our hybrid Monte Carlo dynamics is certainly different from any real 
dynamics, it is still a small-step evolution, so the system has to pass 
through the free-energy barriers. Hence, we expect relaxation times 
obtained in this way to reflect the actual kinetic properties of the system.

Our simulations of the kinetics have been performed for five different $T$. 
The trajectory length $n\epsilon=0.25$ was held fixed, and the average
acceptance rate was typically 85\% or higher.   
\subsection{Measurements}
\begin{figure}[tbh]
\begin{center}
\vspace{-42mm}
\mbox{\hspace{-31mm}\psfig{figure=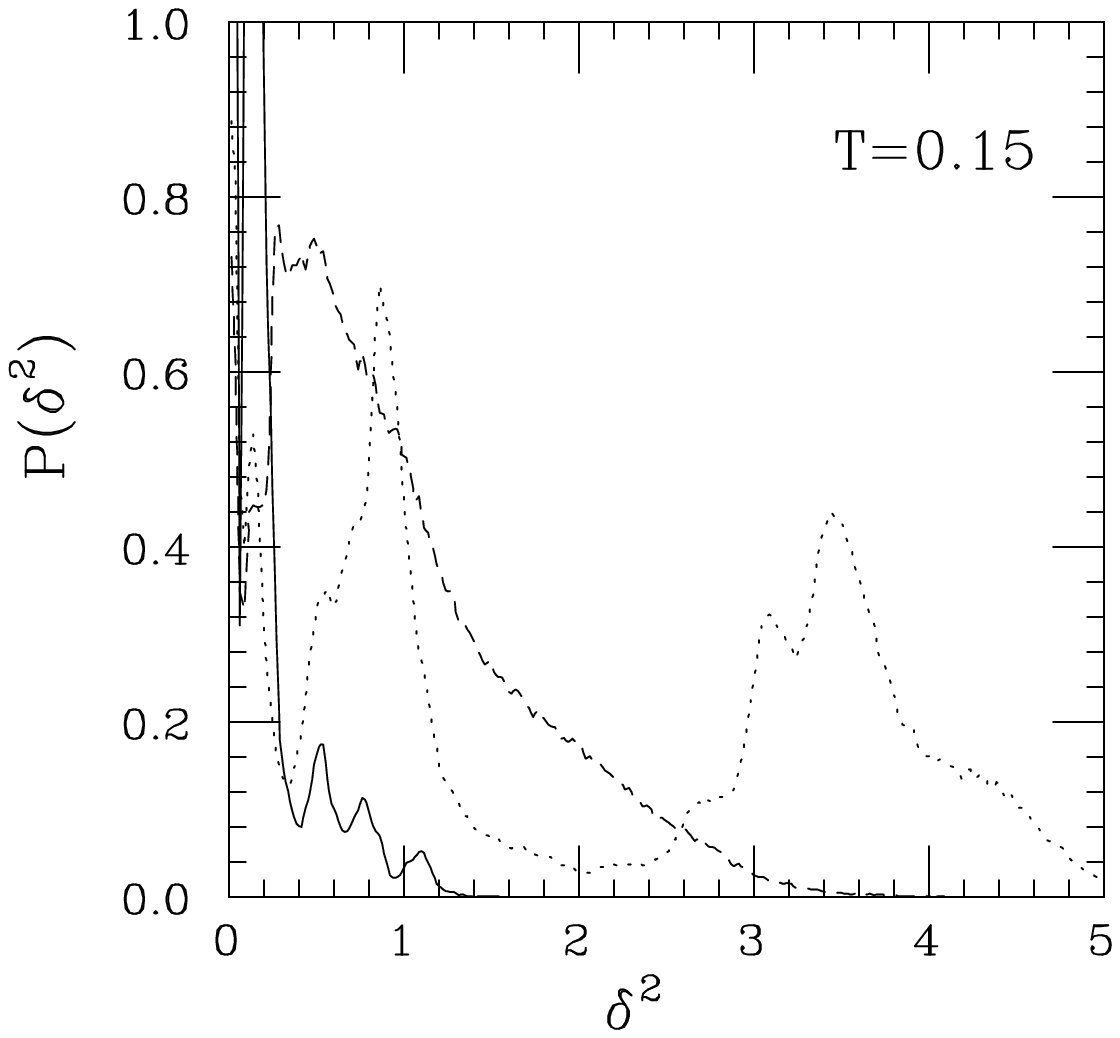,width=10.5cm,height=14cm}
\hspace{-30mm}\psfig{figure=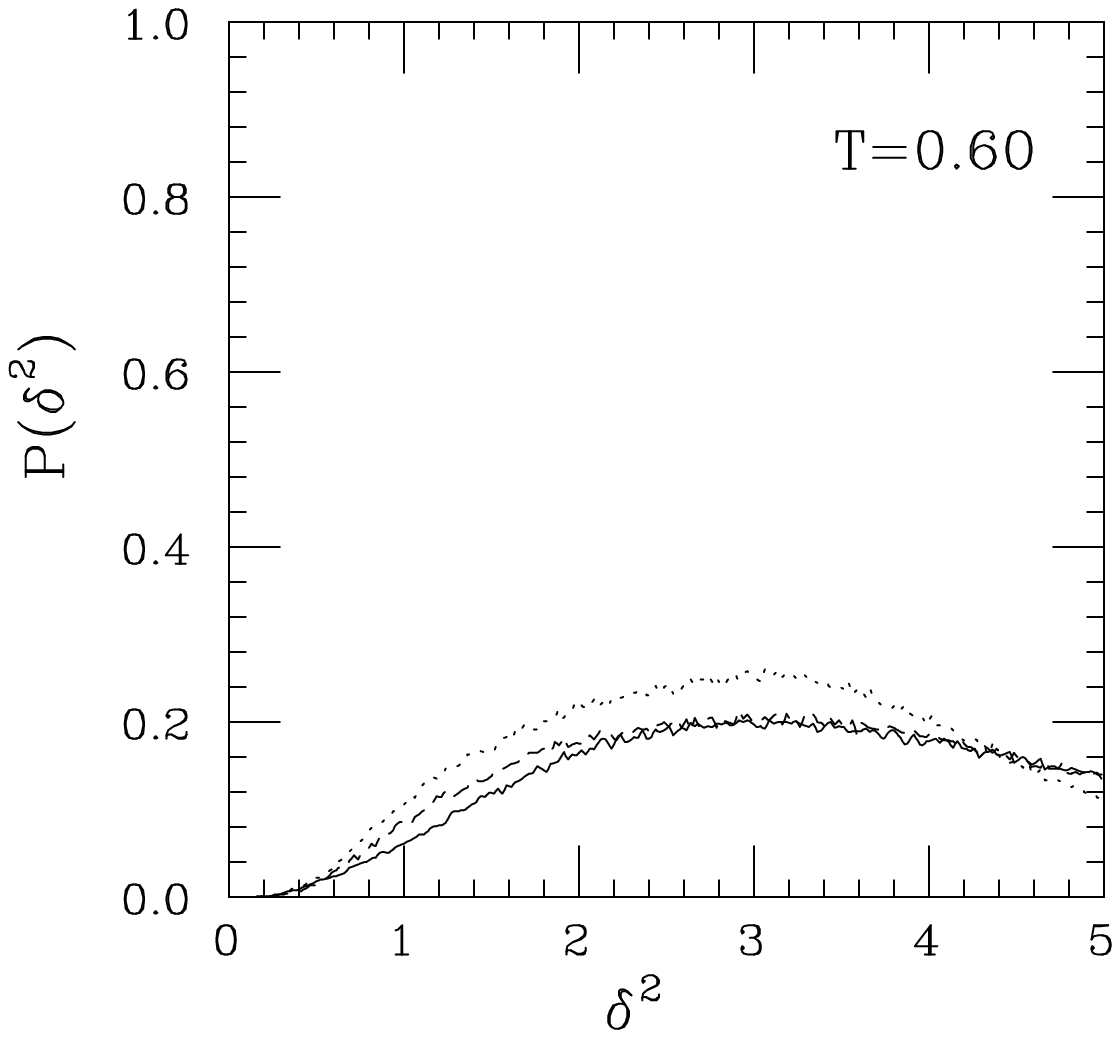,width=10.5cm,height=14cm}}
\vspace{-42mm}
\end{center}
\caption{$P(\delta^2)$ for $T=0.15$ and $T=0.60$ respectively for the sequences 
in Table~1; 81 (solid line), 10 (dashed line) and 50 (dotted line). For $T=0.15$ the 
distribution for sequence 81 is dominated by two narrow peaks at small $\delta^2$ 
which extend outside the figure with a maximum value of around 20.}
\label{fig:1}
\end{figure}

In our thermodynamic simulations the main goal is to find out whether    
the system exists in a state with well-defined shape. 
To address this question, we introduce the usual 
mean-square distance between configurations. For two configurations $a$ and 
$b$ we define
\beq
\delta^2_{ab} = \min {1\over N}\sum_{i=1}^N 
|\bar x^{(a)}_i-\bar x^{(b)}_i|^2
\label{delta}
\eeq
where $|\bar x^{(a)}_i-\bar x^{(b)}_i|$ denotes the distance 
between the sites $\bar x^{(a)}_i$ and $\bar x^{(b)}_i$
($\bar x^{(a)}_i,\bar x^{(b)}_i\in R^2$), 
and where the minimum is taken over translations, rotations, reflections 
and orientations. The probability distribution of $\delta^2$ for fixed 
temperature $T$ and sequence $\sigma$ is given by
\beq
\label{Pdel}
P(\delta^2)= \frac{1}{Z(T,\sigma)^{2}}\int 
d\theta^{(a)}d\theta^{(b)}
\delta(\delta^2-\delta_{ab}^2)
{\rm e}^{-E(\theta^{(a)},\sigma)/T)}
{\rm e}^{-E(\theta^{(b)},\sigma)/T)} ,
\eeq
where $\delta(..)$ denotes the Dirac delta function. $P(\delta^2)$ 
is a very useful quantity~\cite{iori} for describing the magnitude of 
the relevant thermodynamic fluctuations, and can be 
determined numerically. In Fig.~\ref{fig:1} we show three 
examples of $\delta^2$-distribution at two different temperatures.
In what follows we will also frequently use the mean of $P(\delta^2)$,
\beq
\langle \delta^2 \rangle = \int d\delta'^2 P(\delta'^2)\delta'^2
\eeq

\begin{figure}[tbp]
\begin{center}
\vspace{-45mm}
\mbox{\hspace{0mm}\psfig{figure=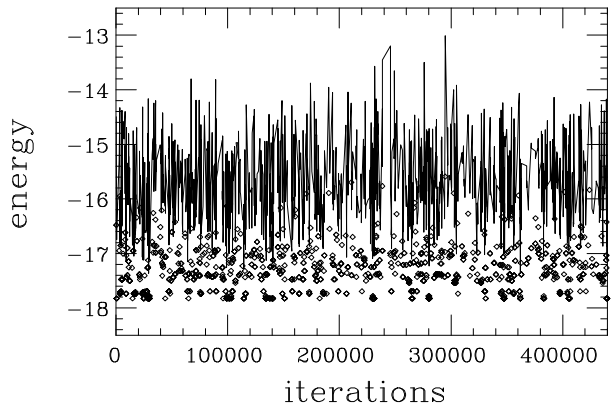,width=10.5cm,height=14cm}}
\vspace{-45mm}
\end{center}
\caption{Evolution of the quenched (diamonds) and unquenched (line) energies 
in the simulation of sequence 10 (see Table~\protect\ref{tab:1}). Measurements were 
taken every 10 iterations. Shown are the data corresponding to the lowest allowed 
temperature.}   
\label{fig:2}
\end{figure}

The energy level spectrum can be studied by using a quenching procedure, 
where whenever the lowest allowed temperature value is visited, the system is 
quenched to zero temperature by means of a conjugate gradient minimization. 
With this method the ground states are found for most of the sequences. One 
reason for believing this is that the two or four symmetry-related copies of 
the lowest-lying minimum were all visited in the simulations. In 
Fig.~\ref{fig:2} we show the evolution of the quenched and unquenched 
energies in one of the simulations. 

\begin{table}[htb]
\begin{center}
\begin{tabular}{|c|c|} \hline
81       & $A A A B A A A B A A A B B A A B B A A A$ \\ 
10       & $A B A A A B B A A B A A A A A A A A B B$ \\ 
50       & $B A B A A A A A A A B A A A A B A A B B$ \\ 
\hline 
\end{tabular}
\caption{Examples of three sequences.}
\label{tab:1}
\end{center}
\end{table}

\section{Thermodynamic and Kinetic Properties}

In this section we present the results of our thermodynamic and
kinetic simulations. Based on these results, we then formulate
criteria for good folding sequences.       

\subsection{Thermodynamics}

The thermodynamic behavior of the $AB$ model has been studied previously for 
chain length $N=8$ and 10~\cite{irback}. This study showed that whether 
or not the chain exhibits a well-defined structure depends strongly on the 
sequence, at fixed temperature. Using the dynamical-parameter method, 
we have now been able to extend these calculations to $N=20$. The results 
obtained closely resemble those for $N=8$ and 10~\cite{irback}; in particular,
they show that the sequence dependence of the thermodynamic behavior remains   
strong for $N=20$. 

\begin{figure}[tbp]
\begin{center}
\vspace{-42mm}
\mbox{\hspace{-31mm}\psfig{figure=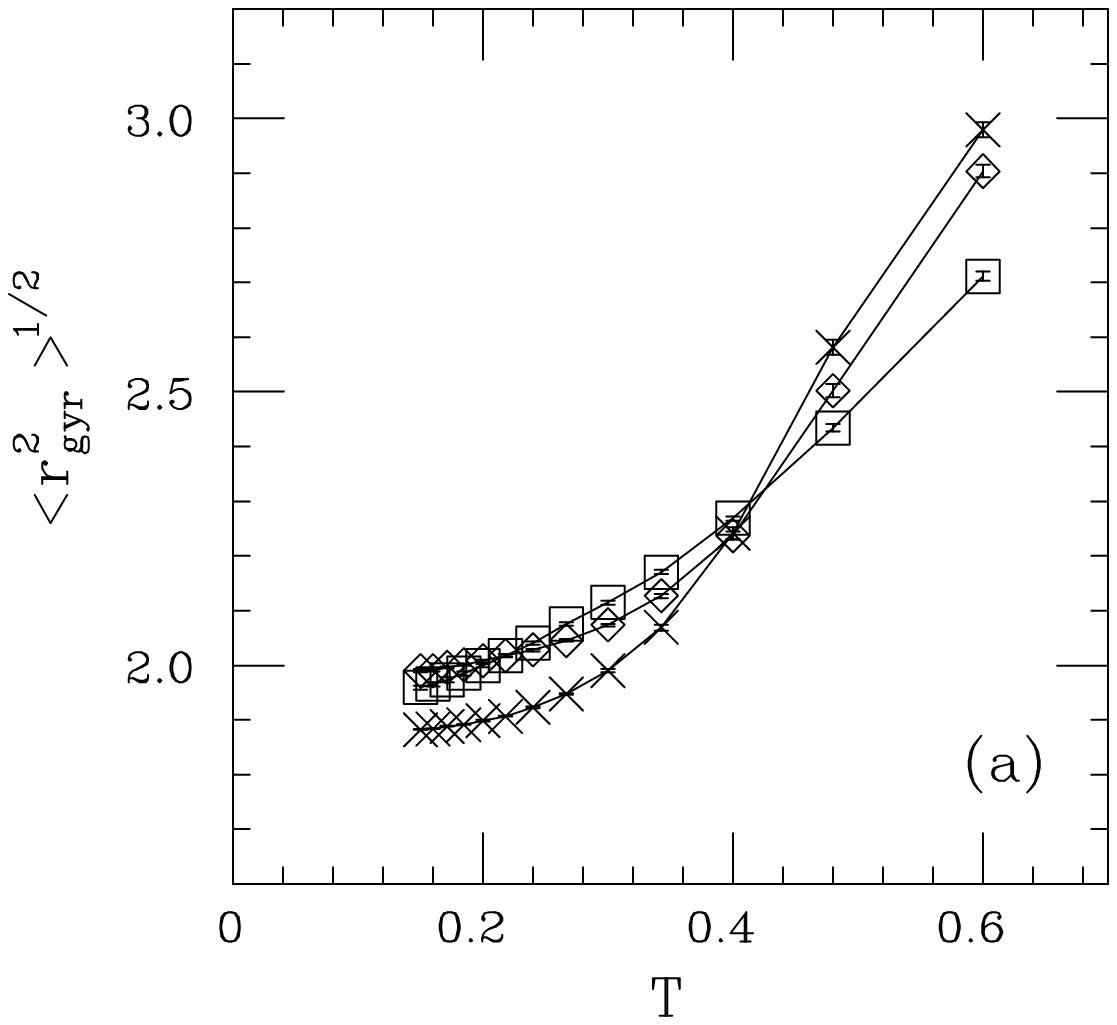,width=10.5cm,height=14cm}
\hspace{-30mm}\psfig{figure=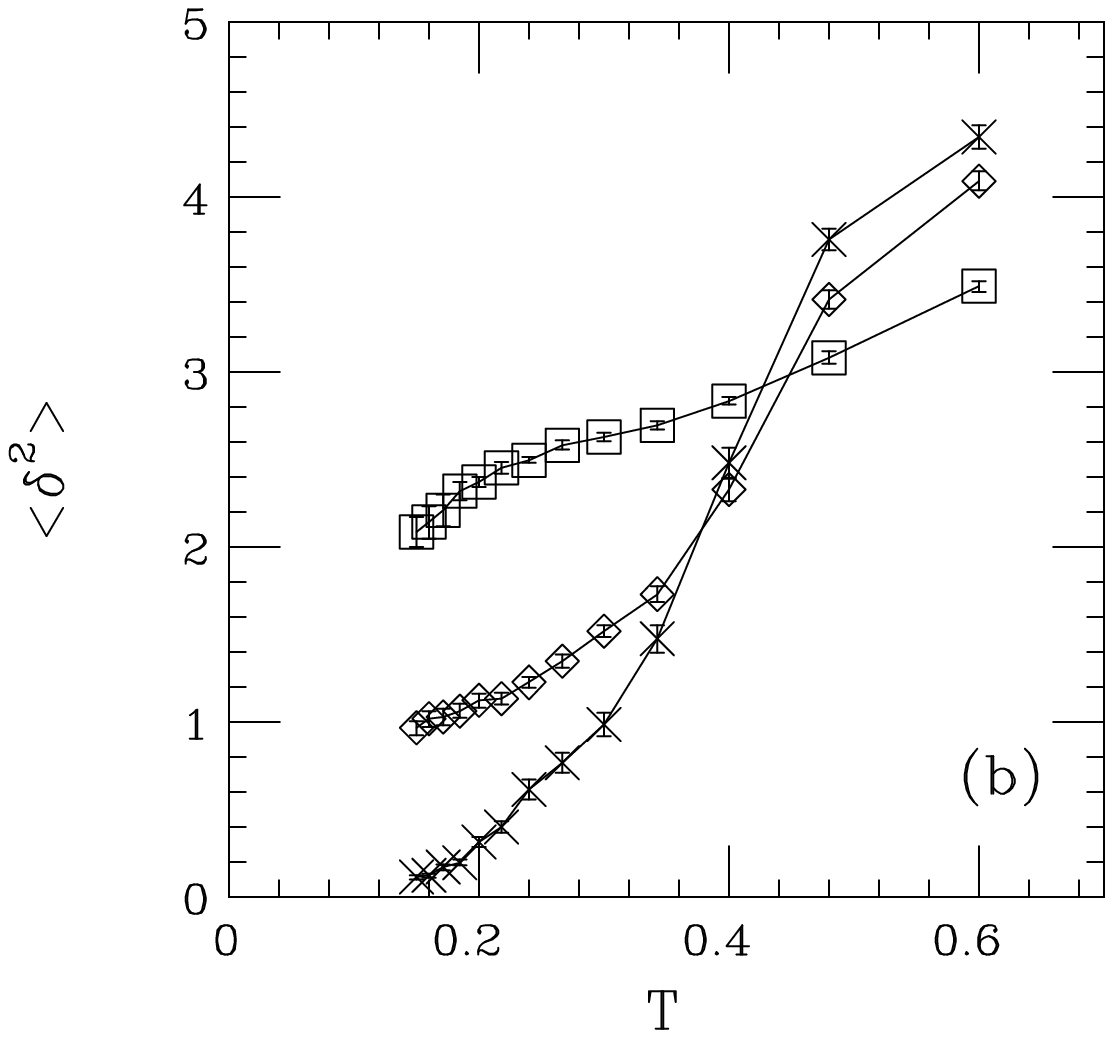,width=10.5cm,height=14cm}}
\vspace{-42mm}
\end{center}
\caption{The temperature dependence of {\bf (a)} the radius of gyration 
$\ev{r^2_{gyr}}^{1/2}$ and {\bf (b)} $\ev{\delta^2}$ for the sequences in 
Table~\protect\ref{tab:1}: 81 ($\times$), 10 ($\diamond$) 
and 50 ($\Box$).}
\label{fig:3}
\end{figure}

In order to illustrate this, we show in Figs.~\ref{fig:1} and \ref{fig:3}
results for the three sequences in Table~1. In Fig.~\ref{fig:1} 
the $\delta^2$ distributions are shown for the lowest and highest 
temperature studied, $T=0.15$ and 0.60. At $T=0.60$ the distributions are 
similar, and show that the fluctuations in shape are large. At $T=0.15$ 
$P(\delta^2)$ is, by contrast, strongly sequence dependent. 
We see that one chain exists in a state with very well-defined shape, 
while the other two still undergo large fluctuations. The differences in 
$P(\delta^2)$ are clearly reflected in the mean value $\ev{\delta^2}$, 
which is shown as a function of $T$ in Fig.~\ref{fig:3}b. 

Although these three chains behave in very different ways, they have similar  
extension. In fact, from Fig.~\ref{fig:3}a it can be seen that the differences in   
radius of gyration are 10\% or smaller at all the temperatures studied. 
Also, we note that the radius of gyration decreases gradually with 
decreasing temperature. No abrupt changes can be seen. 

\subsection{Kinetics}

Next we study the time needed by the system to find the minimum energy 
configuration. Using hybrid Monte Carlo dynamics, we monitor the 
mean-square distance $\delta_0^2$ to this configuration (see Eq.~\ref{delta}). 
The simulations are started from random coils, and as a 
criterion of successful folding we use the condition $\delta_0^2<0.3$. 
At low temperatures the folding time fluctuates widely, which makes the 
average folding time difficult to measure.
For this reason we have chosen to measure 
the probability of successful folding within a given time \cite{sali,socci}.
Following Ref.~\cite{sali}, this quantity will be called the foldicity.
In our simulations the maximum allowed 
folding time is set to 5000 trajectories. For each of the 300 sequences 
we studied five different temperatures, $T=0.15, 0.18, 0.24, 0.34$ and 0.60. 
For each $T$ we carried out 25 simulations for different initial 
configurations.  

The foldicity is expected to be low both at high and low temperature. 
At low temperature the suppression is due to the ruggedness of the 
free-energy landscape. At high temperature folding is slow because the 
search is random.    

\begin{figure}[tbp]
\begin{center}
\vspace{-45mm}
\mbox{\hspace{0mm}\psfig{figure=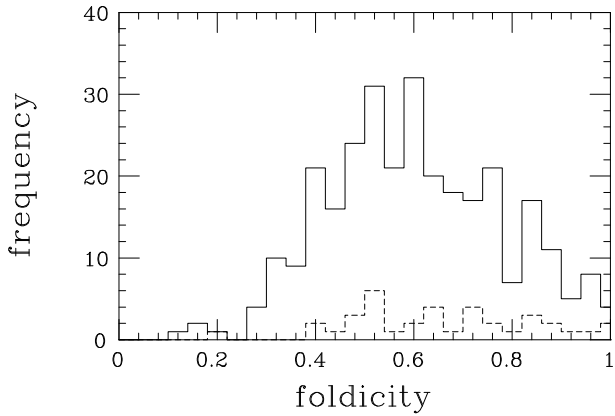,width=10.5cm,height=14cm}}
\vspace{-45mm}
\end{center}
\caption{Histogram of the foldicity at $T=0.34$ for the full set 
of 300 sequences (solid line) and for a subset of 37 sequences that satisfy 
the thermodynamic stability condition in Eq.~\protect\ref{crit} (dashed line).} 
\label{fig:4}
\end{figure}

For a majority of the sequences studied we find that the foldicity 
exhibits a peak in the interval $0.15\le T\le0.60$. 
In order to get precise estimates of the height and location of the 
peak, one would clearly need more data points. However, the available data 
demonstrate that the position of the peak is fairly sequence independent. In fact, 
for 243 of the 300 sequences we obtained a higher foldicity at $T=0.34$ 
than at the other four  temperatures. Also, we note that all sequences have a 
foldicity of 12\% or higher at $T=0.34$, as can be seen from Fig.~\ref{fig:4}. 
Therefore,  it appears that the kinetic behavior has a relatively weak sequence 
dependence.    

\begin{figure}[tbp]
\begin{center}
\vspace{-42mm}
\mbox{\hspace{-31mm}\psfig{figure=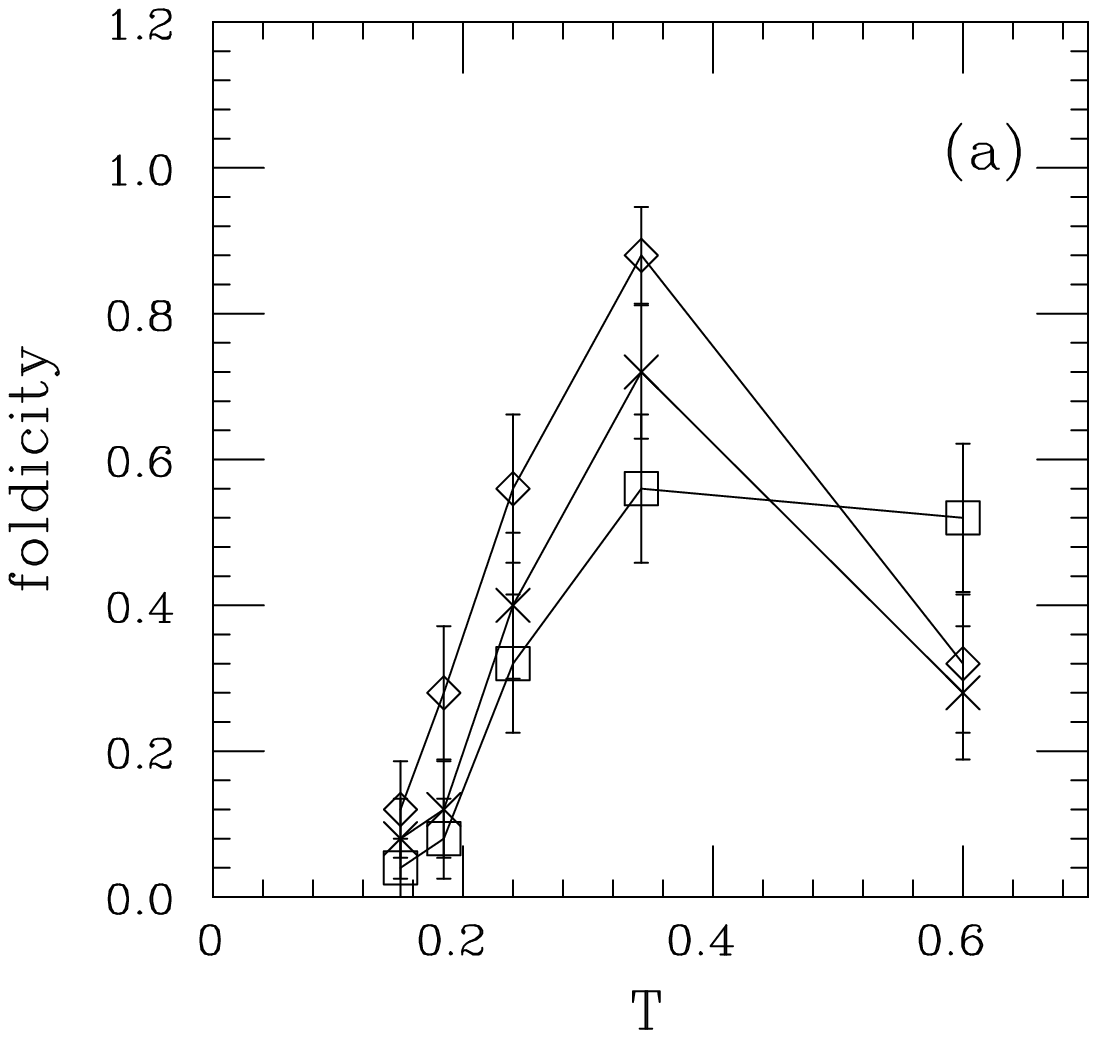,width=10.5cm,height=14cm}
\hspace{-30mm}\psfig{figure=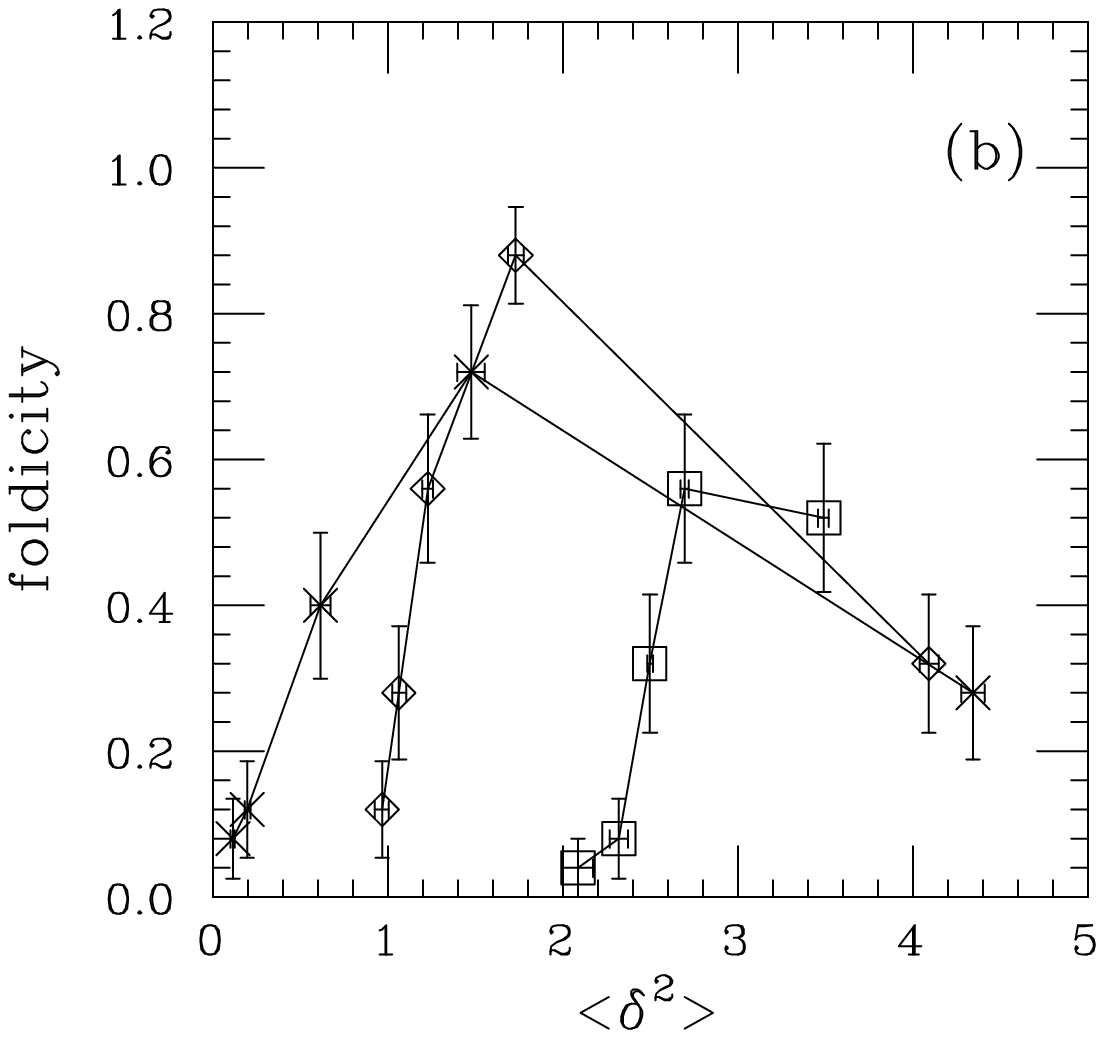,width=10.5cm,height=14cm}}
\vspace{-42mm}
\end{center}
\caption{Foldicity against {\bf (a)} T and {\bf (b)} $\ev{\delta^2}$ for the 
sequences 81 ($\times$), 10 ($\diamond$) and 50 ($\Box$), see 
Table~\protect\ref{tab:1}.}
\label{fig:5}
\end{figure}

In order to illustrate the implications of this, we have in Fig.~\ref{fig:5} 
plotted the foldicity in two different ways. In Fig.~\ref{fig:5}a foldicity  is
plotted against temperature, and in  Fig.~\ref{fig:5}b it is plotted 
against $\ev{\delta^2}$. 
Again, we use the three sequences in Table~\ref{tab:1} as examples.
From a) it can be seen that, at a given temperature, 
the foldicity is roughly similar for the three sequences.
Nevertheless, it follows from b) that their folding properties are 
very different; sequence 10 has a well-defined shape at the point     
where the system freezes, which is not true for the other two 
sequences.  

\subsection{Folding Criteria}

A good folder is a sequence which, at some value of the parameter $T$, exists 
in a unique and kinetically accessible state with well-defined shape.
One way to find out whether or not a given sequence meets this requirement 
is to first locate the lowest temperature at which folding is fast,  
$\tilde T$, and then study the thermodynamic behavior at this 
temperature. If the sequence is found to exhibit a well-defined structure 
at $\tilde T$, then it is a good folder. Alternatively, this can be formulated 
in terms of the folding temperature $T_f$, defined as the temperature where 
the dominance of a single state sets in; a good folder is a sequence with
$T_f>\tilde T$.

In our determination of good folders we take the kinetic quantity $\tilde T$ 
to be same for all sequences. This means that our classification is entirely 
determined by the thermodynamic behavior at a fixed temperature, 
which we take as $\tilde T=0.15$ (cf Fig.~\ref{fig:5}a).
Our motivation for this simplifying approximation is that the kinetic 
behavior has a relatively weak sequence 
dependence, as discussed in the previous subsection. 

With this approximation, a natural criterion for good folders would be to 
require that $\ev{\delta^2}<\tilde\delta^2$ for some $\tilde\delta$. 
Such a cut is appropriate for most sequences. However, some care 
has to be exercised since one might encounter situations where $P(\delta^2)$ 
has a tiny but distant outlier bump, which can make $\ev{\delta^2}$ 
large even though the system spends a large fraction of the time very 
near one particular configuration. Taking this into account we define a 
sequence to be a good folder if
\beq
\ev{\delta^2}<\tilde\delta^2 \qquad {\rm or}\qquad
P(\delta^2<0.1)=\int_0^{0.1}d\delta'^2P(\delta'^2)>\tilde P
\label{crit}\eeq
with $\tilde \delta^2=0.3$ and $\tilde P=0.35$.

With this choice of parameters, we find that 24 of our sequences satisfy
$\ev{\delta^2}<\tilde\delta^2$ whereas 30 satisfy $P(\delta^2<0.1)>\tilde P$.
Our set of good folders, which satisfy one or both of these conditions,
contains 37 of the 300 sequences, or 12\%.

The precise number of sequences classified as good folders depends, of course,
on the choice of $\tilde \delta$ and $\tilde P$, which to some extent is 
arbitrary. However, it should be stressed that for most of the sequences
the classification is unambigiuous in the sense that it is insensitive to
small changes of $\tilde \delta$ and $\tilde P$. Furthermore, we note that
the foldicity distribution for good folders is similar to that for  
all sequences, as can be seen from Fig.~\ref{fig:4}. If these distributions
had been different, the use of a sequence independent $\tilde T$
would have been unjustified.

\section{Sequence Characteristics of Good Folders}

In this section we analyze how $\ev{\delta^2}$ depends upon the binary patterns 
of the sequences.  This is done in two steps. First we make a statistical analysis 
in terms of correlations.  Second, we employ a feedforward Artificial Neural Network 
(ANN) to predict $\ev{\delta^2}$  given the sequence as input. 

\subsection{Choice of Variables}

It turns out to be enlightening and profitable  to transform the original 
binary patterns into more global variables prior to performing the statistical and ANN 
analysis. The following variables are formed:

{\bf Random Walk Representations -- $r_n$}. 
In order to build in some long range correlation properties we consider 
random walk representations  
\beq
\label{rw}
r_n=\sum_{i=1}^n\sigma_i \qquad n=1,\ldots,N
\eeq

{\bf Block Fluctuations -- $\fl_i$ and $\fl$.}
For a block size $s$ we define the variables \cite{irback1} 
\beq 
\bl_i=\sum_{j=1}^s\sigma_{(i-1)s+j}=r_{is}-r_{(i-1)s}
\qquad i=1,\ldots,N/s 
\label{block}
\eeq
In order to efficiently capture the fluctuations of the block variables we introduce the
normalized variables
\beq
\fl_i={1\over K}
\Bigl(\bl_i-{s\over N}\sum_{j=1}^{N/s}\bl_j\Bigr)^2\qquad i=1,\ldots,N/s
\label{fluct}
\eeq
and the (normalized) mean-square fluctuation of the block variables
\beq
\fl={s\over N}\sum_{i=1}^{N/s}\fl_i
\label{msfluct}
\eeq
where the constant $K$ can be found in Ref. \cite{irback1}

{\bf Number of Internal $A$'s --  $N_{IA}$.}
This is the number of $\sigma$=+1 surrounded on both sides by $\sigma$=+1. 
For boundary residues, a single adjacent $\sigma$=+$1$ is sufficient for 
giving a count. The reason for this choice of variable is that in the 
homopolymer limit, $AAAA...A$, the energy landscape is degenerate  
and, hence, the fluctuations are large. Therefore, one expects  
long stretches of $A$'s (or $B$'s) to be rare in good folders, and that 
$N_{IA}$ tends to be low for such sequences. 

{\bf  Number of Clumps -- $N_C$.}
This quantity is defined as the number of clumps 
of $\sigma$=$\pm 1$. The reason for including this variable 
is similar to what was argued for $N_{IA}$ above; it seems natural 
to expect a high $N_C$ for good folders.  

We expect these preprocessed variables, which of course are not independent of 
each other, to shed more light on the structure than the ``raw'' $\sigma$=$\pm 1$ ones, 
when it comes to relate the sequences to $\ev{\delta^2}$.

In this section we will make use of all patterns, even those 
generated by symmetry giving rise to the same $\ev{\delta^2}$. 
As mentioned in Sec.~2.2, 4 of the 300 original sequences are  
symmetric, which means that we have a total of 596 patterns at our 
disposal. 

\subsection{Correlations}

\begin{figure}[tbp]
\begin{center}
\vspace{-45mm}
\mbox{\hspace{0mm}\psfig{figure=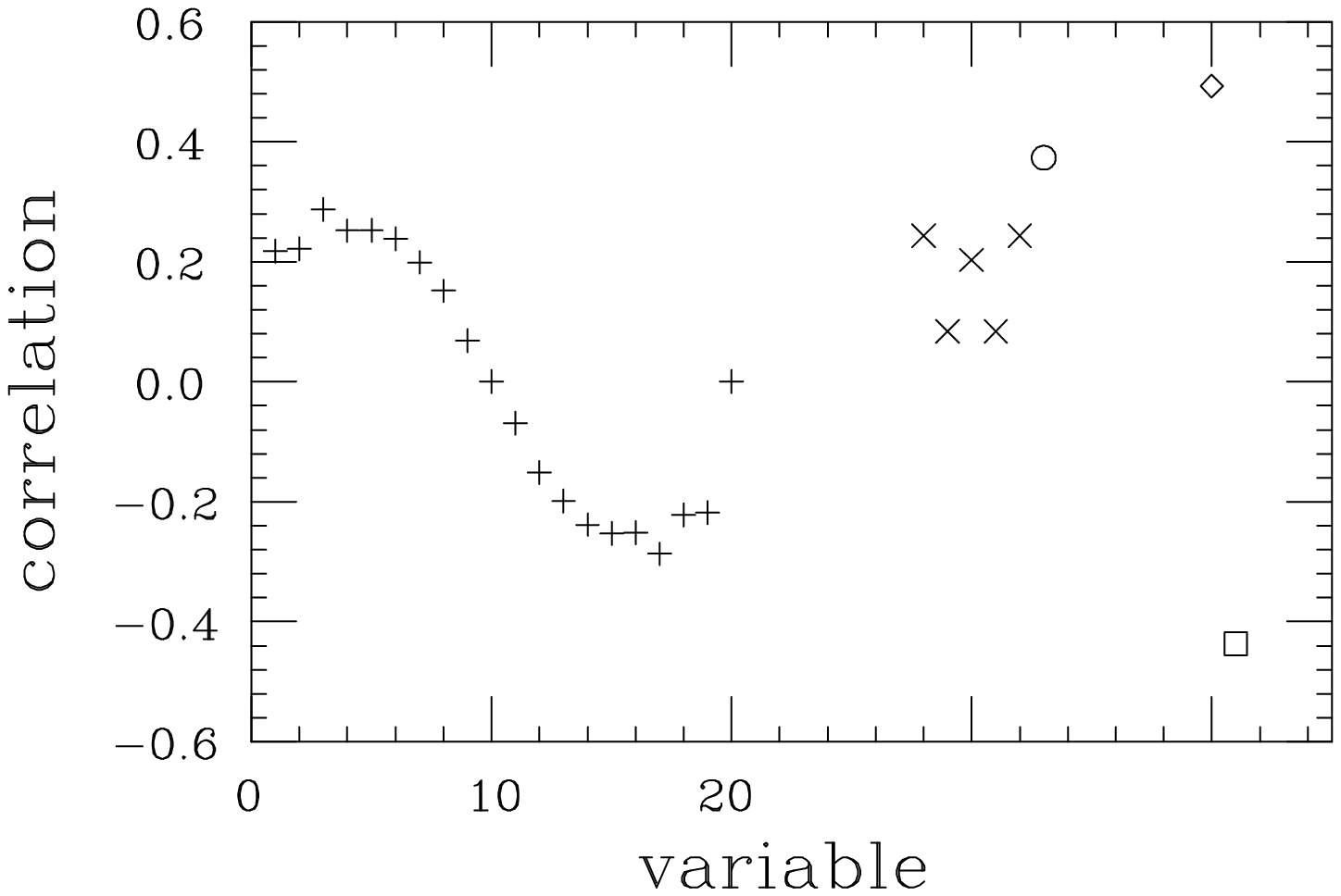,width=10.5cm,height=14cm}}
\vspace{-45mm}
\end{center}
\caption{Correlations of $\ev{\delta^2}$ against  $r_n$ [$n$=1,\ldots,20] ($+$), 
$\ff_i$  [$i$=1,\ldots,5] ($\times$),  $\ff$ ($\circ$), $N_{IA}$ ($\diamond $) 
and $N_C$ ($\Box $).} 
\label{corr}
\end{figure}

To what extent is $\ev{\delta^2}$ correlated with $r_n$, $\ff_i$, $\ff$, $N_{IA}$ and $N_C$?  
In Fig.~\ref{corr} the correlation between  
$\ev{\delta^2}$ and these variables are shown.

As can be seen from Fig.~\ref{corr} the correlations are substantial. It should 
also be mentioned that the same correlation patterns emerge when reducing the data 
set by a factor four. The symmetry for $r_n$ around $n$=10 is inherent from the 
way we have generated the data, and leaves us with nine independent measurements,  
corresponding to e.g. $r_1$ through $r_9$.
The following general observations can be made:
\begin{itemize}
\item $\ev{\delta^2}$ and $r_n$ exhibit sizable positive correlations near the 
endpoints, implying that proteins with many $A$'s near the ends do not fold well.
\item $\ev{\delta^2}$ and the block fluctuations, $\ff_i$ and $\ff$, also show 
strong positive correlations. This is consistent with what was observed for real 
proteins with limited net hydrophobicity in Ref.~\cite{irback1}, where these 
variables are anticorrelated as compared to what is expected from random 
hydrophobicity distributions.
\item The strongest positive correlation is between $\ev{\delta^2}$ and  the 
number of internal $A$'s ($N_{IA}$). This is in line with what is expected 
according to the motivation when introducing $N_{IA}$ above.
\item Related to positive correlation of $N_{IA}$ is the strong anticorrelation 
between $\ev{\delta^2}$ and the number of clumps $N_C$.
 
\end{itemize}
\subsection{Artificial Neural Networks}

Given the substantial correlations between $\ev{\delta^2}$ and the various 
quantities formed out of the sequence patterns, it should be possible to make 
a regression model. If enough data are available for an efficient and reliable fit 
one should be able to predict the folding properties given a binary sequence. 
The state-of-the-art technique for such modeling are feedforward ANN 
(see e.g. Ref.~\cite{hertz}), which will be used here. This method has the 
advantage of capturing non-linear dependencies in a generic way in contrast 
to standard linear regression approaches.  In our case the feedforward ANN 
consists of an input layer representing the variables defined in Sec.~5.1 above, 
an output unit for $\ev{\delta^2}$ and a set of hidden units in order to model 
nonlinearities. The weights connecting the nodes are the parameters of the system. 
In order to avoid overfitting, the number of weights (parameters) should be less 
than the number of ``training'' patterns. Also, some of the patterns should be set aside 
for ``testing''.  Using all the quantities (only $r_1$ through $r_9$ due to symmetries) 
defined in Sec.~5.1 with 5 hidden units implies  17 $\cdot$ 5 + 5 = 90 parameters.

\begin{figure}[tbp]
\begin{center}
\vspace{-45mm}
\mbox{\hspace{0mm}\psfig{figure=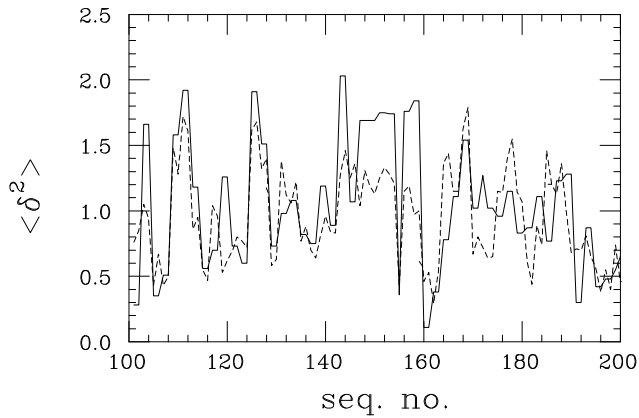,width=10.5cm,height=14cm}}
\vspace{-45mm}
\end{center}
\caption{Comparison of predicted and true values, $\ev{\hat{\delta}^2}$ and 
$\ev{\delta^2}$ respectively,  using $K$=3 for sequences 100 to 200. 
In order to guide the eye, lines connecting the points are drawn. The full line 
represents $\ev{\delta^2}$ and the dashed line $\ev{\hat{\delta}^2}$. } 
\label{res}
\end{figure}

The network was trained using the {\tt JETNET 3.0} package \cite{jetnet} with 
an initial learning rate $\eta_0$=0.5, which decreases  according to 
$\eta_k =0.998 \eta_{k-1}$ and momentum $\alpha$=0.7. In order to obtain 
as reliable performance as possible, the  $K$-fold cross validation procedure was 
used, where the data set was randomly divided into $K$ equal parts. Each of the 
the $K$ different parts were once used as test sets, while the remaining $K-1$ 
sets were used for training. In our problem sets with  $K$=2, 3, and 4 were used. 
In Fig.~\ref{res} the resulting prediction, $\ev{\hat{\delta}^2}$, is compared with 
the true values, $\ev{\delta^2}$,  for a representative subset of the 
sequences using $K$=3 is shown. 
As can be seen from Fig.~\ref{res} the predicted values follow the true values 
pretty well even though sometimes quantitative agreement is lacking. 
The degree of success can be quantified using the standard error of prediction, 
\beq
R=\frac{1}{\sigma^2}\frac{1}{N}\sum ( \ev{\hat{\delta^2}} -\ev{\delta^2})^2
\eeq
where $\sigma$ is the standard deviation of $\ev{\delta^2}$ for all 596 patterns.
For $K$=3 and 4 we find $R$=0.69$\pm$0.02. This value deteriorates somewhat 
if a smaller subset of the data is used, which indicates that with a larger available 
data set  the performance is likely to improve. In the long term one would like 
to predict a classification (folder/non-folder) but the limited folding data at our 
disposal do not yet allow for that.

\section{Non-Randomness  and Folding}

In Sec.~5 we studied the difference between folding and 
nonfolding sequences in the $AB$ model. This could be done in a
controlled way due to the fact that we have unbiased samples of
folding and nonfolding sequences at our disposal. To assess the
applicability of our methods to real amino acid sequences is
more difficult, and we shall not deal here with the problem of  
predicting the behavior of individual amino acid sequences. However, 
we would like to stress that the binary hydrophobicity patterns corresponding 
to a large class of functional proteins do exhibit interesting similarities 
with folding sequences in the $AB$ model, as was demonstrated in 
Ref.~\cite{irback1}. The proteins sequences considered in  Ref.~\cite{irback1} 
have a limited net hydrophobicity
\beq
X={N_+-Np\over \sqrt{Np(1-p)}}
\label{X}
\eeq
where  $N_+$ is the number of hydrophobic residues, $N$ is the
total number of residues, and $p$ is the average of $N_+/N$ over
all sequences. In the following,
we consider sequences with $|X|<0.5$. 

\begin{figure}[tbp]
\begin{center}
\vspace{-42mm}
\mbox{\hspace{-31mm}\psfig{figure=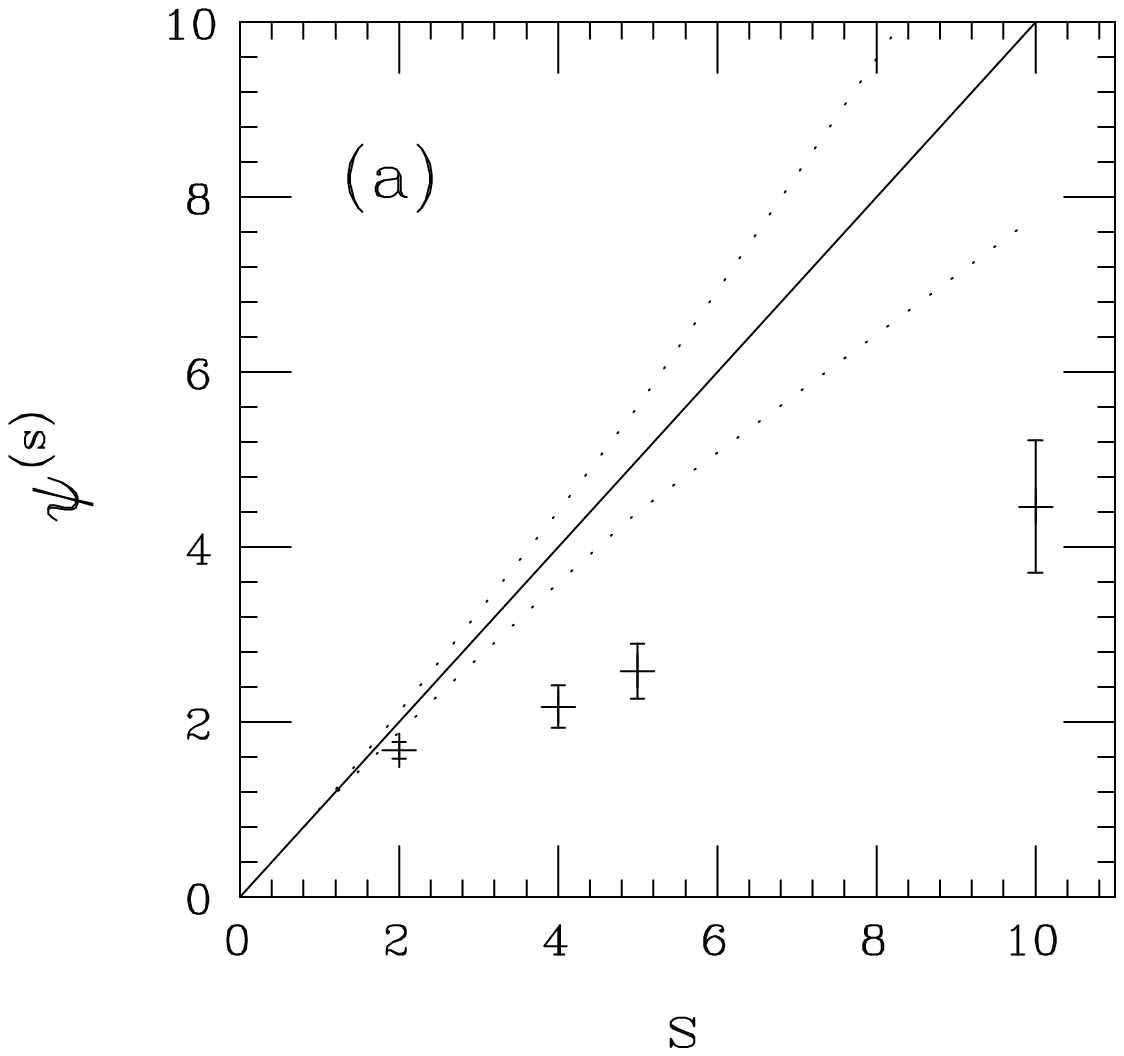,width=10.5cm,height=14cm}
\hspace{-30mm}\psfig{figure=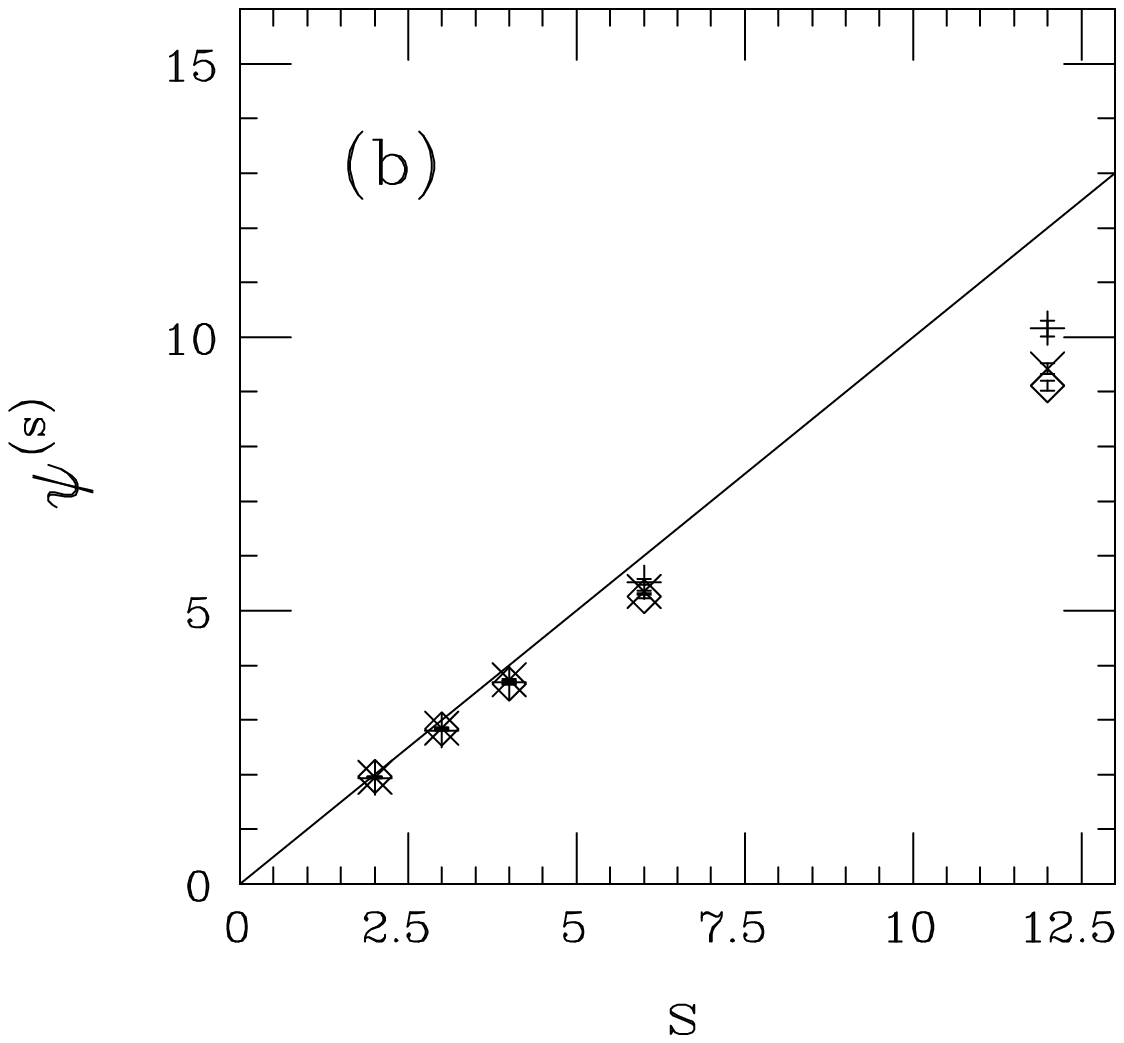,width=10.5cm,height=14cm}}
\vspace{-42mm}
\end{center}
\caption{Mean-square fluctuation of the block variables, $\fl$, against block 
size $s$. 
{\bf (a)} Good folding sequences in the $AB$ model. Also shown are
the mean $s$ (full line) and the $s\pm\sigma$ band (bounded by dotted lines) 
for random sequences \protect\cite{irback1}.
{\bf (b)} Functional proteins for $|X|<0.5$, $50<N\le 150$ 
(+; 2457 qualifying proteins), 
$150<N\le 250$ ($\times$; 2228), and $250<N\le 350$ ($\diamond$; 1642). 
All data are from the SWISS-PROT data base \protect\cite{swiss-prot}.
The straight line is the result for random sequences \protect\cite{irback1}.}   
\label{fig:8}
\end{figure}

In order to examine these two groups of binary strings, it is instructive     
to consider the scaling of the fluctuation variable $\fl$ with block 
size $s$~\cite{irback1}. In Fig.~\ref{fig:8} we show results for this
quantity. For both groups, it can be seen that 
$\fl$ grows significantly slower with $s$ than for random sequences.
This behavior implies that the sequence variables $\sigma_i$  
exhibit anticorrelations. A simple way to see that is to consider 
the one-dimensional Ising model, where each configuration is 
given a statistical weight
\beq
P\propto\exp\Bigl(K\sum_{i=1}^{N-1}\sigma_i\sigma_{i+1}\Bigr)
\label{ising}
\eeq
For $K=0$ the $\sigma_i$'s are completely random, while $K>0$ and $K<0$ 
correspond to ferromagnetic and antiferromagnetic behavior respectively.
It turns out that one obtains results similar to those seen in 
Figs.~\ref{fig:8}a and b for $K=-0.25$.

\section{Summary}

Fairly detailed studies of the folding properties of a toy model for proteins, 
containing effective hydrophobicity interactions only, have been performed. 
The thermodynamic quantity $\ev{\delta^2}$  exhibits a strong sequence dependence 
in contrast to foldicity, which measure the kinetic properties. Hence criteria for 
chains with good folding properties have been devised solely in terms of $\ev{\delta^2}$. 
With these criteria,  approximately 10\% of the 300 generated sequences are 
classified as good folders. These conclusions have been possible due to extensive 
searches using an efficient dynamical-parameter algorithm, which 
with very large probability visits the ground states. 
A similar fraction of good folders was obtained in the lattice model
study of Ref.~\cite{sali}, where 30 out of 200 randomly chosen sequences 
were classified as good folders.

Our conclusion that the thermodynamic properties are more important 
for the classification of folders/non-folders than the kinetic ones is
in line with the lattice model results of Ref.~\cite{bryngelson}. 
These authors introduced a kinetic glass transition temperature, 
which was found to be nearly sequence independent. Although the kinetic 
studies of Ref.~\cite{bryngelson} are more elaborate than ours, it should
be stressed, however, that our data provide no justification for introducing
this transition temperature, which may indicate a difference between 
the models.

Using statistical and artificial neural network methods, substantial functional 
dependencies between sequence patterns and $\ev{\delta^2}$ are revealed. With 
larger statistics it should be possible given the sequence pattern to predict 
$\ev{\delta^2}$ within a reasonable confidence level; in other words to predict 
whether a given sequence folds or not.

Related to this strong correlation is the observed pattern of the 
non-randomness for the folders, which show similar qualitative behavior  
with what is observed for real proteins \cite{irback1}.

\end{document}